\documentstyle[12pt]{article}
\newcommand{\beq}{\begin{equation}}
\newcommand{\eeq}{\end{equation}}

\def\half{{\textstyle{1\over2}}}

\def\quart{{\textstyle{1\over4}}}

\def\p1half{{\textstyle{{{p+1}\over{2}}}}}

\def\23phalf{{\textstyle{{{23-p}\over{2}}}}}

    \let\p=\pi

 \def\bd{\begin{document}} \def\ed{\end{document}}
\def\ds{\documentstyle} \let\fr=\frac \let\bl=\bigl \let\br=\bigr
\let\Br=\Bigr \let\Bl=\Bigl
\let\bm=\bibitem
\let\na=\nabla
\let\pa=\partial \let\ov=\overline
\newcommand{\be}{\begin{equation}}
\newcommand{\ee}{\end{equation}}
\def\ba{\begin{array}}
\def\ea{\end{array}}
\def\ft#1#2{{\textstyle{{\scriptstyle #1}\over {\scriptstyle #2}}}}
\def\fft#1#2{{#1 \over #2}}
\def\del{\partial}
\def\sst#1{{\scriptscriptstyle #1}}
\def\oneone{\rlap 1\mkern4mu{\rm l}}
\def\ie{{\it i.e.\ }}
\begin{document}
\thispagestyle{empty}
\begin{titlepage}

\bigskip
\hskip 3.7in{\vbox{\baselineskip12pt
}}

\bigskip\bigskip
\centerline{\large\bf Electric-Magnetic Duality, Matrices, \& Emergent Spacetime}

\bigskip\bigskip
\bigskip\bigskip
\centerline{\bf Shyamoli Chaudhuri\footnote{Based on talks
given at the {\em Workshop on Groups \& Algebras in 
M Theory}, Rutgers University, May 31--Jun 04, 2005. Email: shyamolic@yahoo.com}}
\bigskip
\centerline{1312 Oak Drive}
\centerline{Blacksburg, VA 24060}
\bigskip\bigskip

\bigskip
\begin{abstract}
This is a rough transcript of talks given at the {\em Workshop on Groups
\& Algebras in M Theory} at Rutgers University, May 31--Jun 04, 2005.
We review the basic motivation for a pre-geometric formulation of
nonperturbative String/M theory, and for an underlying eleven-dimensional
electric-magnetic duality, based on our current 
understanding of the String/M Duality
Web.
We explain the concept of an {\em emerging   
spacetime geometry} in the large $N$ limit of a $U(N)$ flavor 
matrix Lagrangian, distinguishing our proposal from generic proposals 
for quantum geometry, and explaining why it can incorporate curved
spacetime backgrounds. 
We assess the 
significance of the extended symmetry algebra of the matrix Lagrangian, 
raising the question of whether our goal should be a  
duality covariant, or merely duality invariant, Lagrangian. We explain 
the conjectured isomorphism between the $O(1/N)$ 
corrections in any given large $N$ scaling limit of the
matrix Lagrangian, and the corresponding $\alpha^{\prime}$ corrections in a 
string effective Lagrangian describing some weak-coupling
limit of the String/M Duality Web.
\end{abstract}

\end{titlepage}

\section{Introduction}

\vskip 0.1in Understanding the symmetry principles and the fundamental
degrees of freedom in terms of which nonperturbative String/M theory 
is formulated is a problem of outstanding importance in theoretical
high energy physics. The Rutgers Mathematics workshop on {\em Groups \& Algebras
in M Theory} this summer devoted part of its schedule to an assessment of the 
significance of Lorentzian Kac-Moody algebras to recent conjectures
for the symmetry algebra of String/M theory. The status, and future prospects, for 
developments in the representation theory of generic infinite-dimensional 
Lorentzian Kac-Moody algebras also received
intense discussion. My talks at the workshop were devoted to an introductory 
survey of the String/M Duality web, followed by my own proposal for nonperturbative string/M
theory \cite{mat1,mtheory}: {\em a fundamental theory of emergent spacetime}, based on a 
matrix Lagrangian with U(N) flavor symmetry, but also distinguished by extended 
global symmetries, which has the pleasing consequence of yielding the different 
weak coupling limits of the String/M Duality Web in its myriad choices of 
multiple-scaled large N limit. We will enlarge upon this succinct, but weighty, 
one--line summary of our
proposal in this transcript. Our emphasis will be on the worldsheet evidence for an 11D 
electric-magnetic duality in nonperturbative String/M theory \cite{flux,hodge}, and on
clarifying the precise role played by extended global symmetries in our matrix theory 
formulation. Related technical details can be 
found in the references \cite{mat1,mtheory,hidden}.

\vskip 0.1in
Perturbative string theory in flat spacetime 
backgrounds, including possible two-form background
fields, is known to be both renormalizable, anomaly-free, and ultraviolet-finite. 
We use the term {\em perturbatively 
renormalizable}  to describe the target spacetime string theory Lagrangian 
despite the presence of  infinitely many couplings in the $\alpha^{\prime}$ expansion, 
because {\em only a finite number of independent parameters} go into their determination, 
and these are all found at the lowest orders in the $\alpha^{\prime}$ expansion.
The existence of only a finite number of independently renormalized couplings
is the defining criterion for the Wilsonian renormalizability of a quantum theory.
Thus, from this 
perspective, the perturbative string theoretic unification of gravity and Yang-Mills gauge
theories with chiral matter can be seen as providing a precise, and unique, 
gravitational extension of the 
anomaly-free and renormalizable Standard Model of Particle Physics. 

\vskip 0.1in The ultraviolet cutoff, $m_s$ $=$ $\alpha^{\prime -1/2}$, can 
therefore be
taken to infinity while keeping the mathematical framework of 
weakly-coupled perturbative string theory  
reliable, down to arbitrarily short distances. There
is no evidence of quantum corrections to the flat spacetime background
geometry; the so-called 
semi-classical ground state is also the quantum ground state. Assuming that 
the supersymmetry breaking scale in
Nature is far below the string scale, we then 
appear to have a perfectly good perturbative
unification of 
quantum gravity and the supersymmetric Standard Model of astro-particle physics
and cosmology. Note that all indications to date point to the probable 
{\em weak-coupling unification} 
 in our 4D world \cite{chltalk}. In addition, 
strong-weak coupling dualities have, in many cases, enabled 
us to gain considerable insight into the strong coupling physics tied to 
supersymmetry breaking and other open questions relating to dynamics. 
So why do we need to go beyond?

\vskip 0.1in The motivation for developing a {\em pre-geometric} formulation
for nonperturbative string/M theory 
is not because of any evidence for a breakdown of continuum 
spacetime at short distances, not for any shortcomings of the string
theoretic description, and not wholly because we need to understand
strong coupling and nonperturbative physics better. See my recent 
talk \cite{chltalk} enlarging on these particular aspects of string theory.
The essential reason we need a pre-geometric formulation is that 
perturbative string theory is inherently background
dependent: it violates the essence of Einstein's theory of relativity because it separates
the description of matter and interactions, both of which are strings at weak coupling, 
from the description of the continuum target spacetime in which they exist. In the absence of
a nonperturbative formulation in which both target spacetime geometry, and
matter and interactions, emerge on equal footing, this fundamental gap in our 
understanding would remain even if we had a handle on strong coupling physics. This is the
basic motivation for what we have called {\em a fundamental theory of 
emergent spacetime geometry} \cite{mtheory}. 

\vskip 0.1in It is important to distinguish 
the concept of emergent geometry as introduced by us \cite{mat1,mtheory} from the set of
ideas usually called {\em quantum geometry}. In our notion of a fundamental theory of
emergent
geometry based on zero-dimensional matrices, there is no underlying classical 
continuum manifold upon which we build nontrivial metrics out of quantum gravitational 
effects: loop quantum general relativity, 
branched polymer models, or quantum spacetime foam, all assume a
continuum classical scaffolding \cite{quantg}. Nor is emergent geometry a latticization
of a background spacetime geometry.\footnote{Lattice gauge theory does offer
an important analogy in one respect. Ken Wilson's latticization of Yang-Mills gauge theory
was remarkably insightful in that it implemented {\em lattice} gauge invariance: it is a 
gauge invariant nonperturbative 
formulation, even away from the continuum limit. Our matrix Lagrangian achieves the 
same for string theory: the global symmetries of the 
spacetime Lagrangians of perturbatively renormalizable string theories that emerge in
its large $N$ limits 
are already present in the matrix Lagrangian at finite $N$ \cite{mtheory}. We should
note some novel developments in the latticization of rigid SYM theories
with 16 supercharges \cite{latt}, and with fermionic variables living on both links and sites.} 
We begin with a Lagrangian of {\em zero}
dimensional matrices, and a symmetry algebra extending far beyond the
usual $U(N)$. Spacetime itself, and
the target space Lagrangians of perturbatively renormalizable and anomaly-free 
superstring theories, emerge from the myriad large $N$ limits of this matrix
Lagrangian. Each large $N$ limit 
is a quantum ground state of the fundamental theory, and its global symmetry algebra 
originates in the symmetries of the matrix Lagrangian. From the perspective of 
quantum cosmology \cite{hawk}, the large $N$ ground states of the 
matrix Lagrangian {\em are} the quantum final 
states of the Universe, and each quantum state
comes with its individual set of initial conditions. Within the framework of the
Hartle-Hawking paradigm
for a quantum theory of the Universe \cite{hawk},  we can associate each
such \lq\lq final state", a spacetime string effective Lagrangian in specific target spacetime
background, to the endpoint of a
{\em consistent history} in our theory for the quantum mechanics of finite $N$ 
matrices. Following this epoch, the evolution of physics from the string scale down to TeV scale
energies will be governed by conventional, tried-and-true, spacetime quantum field theoretic, renormalization group methods \cite{chltalk}.

\vskip 0.1in Any viable proposal for
nonperturbative String/M theory must incorporate electric-magnetic duality
symmetries \cite{ss,witdual,polbook}. Let us begin by examining the 
evidence for an {\em eleven dimensional} origin for duality. This is a 
key assumption made in 
most discussions of the Duality Web of String/M theory and its global
symmetry algebra, usually also assumed to be the source of the
electric-magnetic dualities that can become manifest in lower spacetime
dimensions.
Electric-magnetic duality in $D$ 
spacetime dimensions interchanges an electrically charged  
$p$brane with a magnetically charged $(d$$-$$p$$-4)$brane, respectively, 
 sources for a pair of dual rank 
$(p$$+$$2)$ and rank ($d$$-$$p$$-2$) field strengths. The ten-dimensional
perturbatively renormalizable and anomaly-free type II superstring 
theories contain a
full spectrum of such pbranes, with $p$ in the range: $-1$ $\le$ 
$p$ $\le$ $9$, covering
D-instantons thru D9branes \cite{dbrane}. This raises the
following puzzle. By a generalization of the Dirac quantization
condition for electric and magnetic point charges in four
spacetime dimensions, one would expect that the product of the
quantum of charge for a Dpbrane, and that of its $d$-dimensional
Poincare dual D(d-p-4)brane, satisfies the relation:
\begin{equation}
\nu_p \nu_{d-4-p} = 2\pi n , \quad \quad n \in {\rm Z}  \quad .
\label{eq:diracq}
\end{equation}
The spectrum of values for $p$ listed above does not cover all of
the expected charges in ten dimensions: we are missing a D(-2)brane and a
D(-3)brane, the 10d Poincare duals of the D8brane and D9brane,
respectively. This is especially puzzling because, in a 
groundbreaking work \cite{dbrane}, Polchinski had shown that the
quantum of Dpbrane charge could be computed from first principles
using the worldsheet formalism of weakly coupled perturbative string theory, 
predicting also that the value of
$n$ in the Dirac quantization relation is {\em unity}. Thus,
although we were expecting to find evidence for electric-magnetic
duality in the full Dpbrane spectrum we appear, instead, to have
found a clash with Poincare-Hodge duality in ten target spacetime
dimensions.

\vskip 0.1in There is a simple
resolution to this puzzle, and it brings with it the key insight that 
the electric-magnetic duality underlying {\em nonperturbative} String/M theory
is inherently eleven dimensional. It is remarkable that the smoking gun for this
insight can be found in a weakly coupled perturbative type IIA
string theory calculation using ordinary worldsheet methods. Having 
thereby discovered concrete evidence for 
the validity of this fundamental symmetry of the nonperturbative 
theory in a hands-on, bottom-up, approach using the perturbative formalism, we
can proceed with confidence to examine its top-down consequences 
starting from elegant first principles. The details of the worldsheet calculation
can be found in the references \cite{cmnp,pair,pairf,flux,hodge}; we will only 
outline the basic result in what follows.

\vskip 0.1in
Consider the anomaly-free and perturbatively renormalizable
weakly coupled 10d type IIA string theory in the
background with 32 D8branes. One of the two supersymmetries of
the IIA string has been broken by the presence of orientifold
planes at $X^9$$=$$0$, and $X^9$$=$$R_9$, and a 
T-duality transformation maps this background to
an analogous background with 32 D9branes of the 10d type IIB
string theory. The stack of coincident Dpbranes lies on a single
orientifold plane, and carries worldvolume Yang-Mills gauge fields 
with gauge group $SO(32)$. At finite string coupling, an eleventh
target space \lq\lq dimension" emerges, corresponding to the vacuum expectation
value of the scalar dilaton field \cite{witdual}. Consider the following
{\em gedanken} experiment in the strongly coupled IIA string theory:\footnote{This
is only a {\em gedanken} experiment at the present time because the strongly coupled 
IIA string theory is, more precisely, M theory compactified 
on an $S^1$$\times$$S^1/{\rm Z}_2$, and we do not know how to calculate
in that 
theory beyond its low energy 11-dimensional supergravity limit.} we 
evaluate the pair correlation function of a pair of spatially
separated Wilson loops wrapping the eleventh dimension. The
limit of pointlike loops, and large spatial separation, corresponds 
to taking $R_{11}$$\to$$0$, and hence we recover a
result in {\em ten-dimensional} type IIA supergravity. Our thought experiment must,
therefore, have a precise analog in the factorization limit of a suitable 
weakly coupled perturbative type IIA string 
amplitude with the worldsheet topology of an annulus.
The relevant computation 
is as follows: consider the Polyakov path integral summing 
over worldsurfaces with the topology of an annulus, and with Dirichlet
boundary conditions on all ten embedding target space dimensions. 
We will require further that the boundaries are 
mapped to a pair of given pointlike loops, ${\cal C}_i$, ${\cal C}_f$,
spatially separated by a distance $R$ in the $X^8$ direction, lying within
the worldvolume of the stack of 32 coincident D8branes on an $O8$ 
plane.  This augmented boundary value problem for embedded
Riemann surfaces with the topology of an annulus 
is the precise supersymmetric type IIA analog of a computation
carried out in 1986 by Cohen, Moore, Nelson, and Polchinski \cite{cmnp}
for the bosonic string theory: the off-shell closed string tree propagator 
between pointlike loops. An analysis of the type IIA macroscopic loop amplitude appears in a 
paper by myself and Novak \cite{pairf}. The factorization limit of this
 amplitude with pointlike loops 
yields the long-range
interaction of a pair of supergravity sources, and the result was found 
by me to be consistent with the tension of a Dirichlet (-2)brane \cite{flux}. 

\vskip 0.1in Notice that the Poincare dual of the D8brane
in ten spacetime dimensions is a D(-2)brane. The field equations, and spacetime Lagrangian,
for Roman's massive type IIA supergravity \cite{romans,berg} do, in fact, include a scalar field strength, 
$F_0$, and the D(-2)brane can be identified as its source. Our worldsheet result 
therefore lays to rest any discrepancy
in 10d electric-magnetic duality if we also accept that there are no consistent 10d
superstring backgrounds with nonvanishing D9brane charge. 
A 9brane would be a supergravity source for an eleven-form field strength, so it is 
clearly not something expected in a ten-dimensional supergravity. From the perspective of
the worldsheet calculation, notice that a Dirichlet boundary with 9+1 worldvolume 
coordinates is special: since the Dirichlet defect fills all of 10d spacetime, there is
nowhere for the emanating \lq\lq flux lines" to go except to probe an eleventh dimension. 
Thus, we must conclude 
that, strictly speaking, a D9brane with nonvanishing flux can only exist in 
the strongly coupled type IIA/M theory
limit since it has an additional, {\em eleventh} embedding target
space dimension. Happily, the magnetic dual of the D9brane in eleven dimensions 
is a D(-2)brane, so that the Dirichlet pbrane spectrum for which we have found evidence
in the worldsheet formulation of the weakly coupled perturbative type II string theories,
namely, $-2$ $\le$ $p$ $\le$ $9$, is consistent with the existence of an eleven 
dimensional electric-magnetic duality in the nonperturbative theory \cite{flux,hodge}. 
It is remarkably fortuitous 
that a remnant of 11D electric-magnetic duality survives in the 
weakly coupled 10d limits. Both the electric D(-2)brane, and its
magnetic D9brane dual, make an appearance in the standard worldsheet formalism
of perturbative superstring theory.

\vskip 0.1in Does 11d electric-magnetic duality imply that nonperturbative
String/M theory is necessarily a theory formulated in 11 target space dimensions?
Surprisingly, the answer turns out to be {\em No}, a point that can be made
from a variety of different perspectives \cite{mat1,w4,hidden,mtheory}. The matrix
formulation for a theory of emergent spacetime proposed by us in \cite{mtheory} 
achieves precisely this objective:
as many as eleven spacetime coordinates can emerge in the large $N$
limits of what starts out as a {\em zero-dimensional} matrix Lagrangian with flavor
$U(N)$ symmetry.
Electric-magnetic duality can be built into this formalism by specifying the 
extended symmetry algebra ${\cal G}_s$$\times$$\rm G$$\times$$U(N)$, 
where $\rm G$ is the electric-magnetic self-dual Yang-Mills gauge group, and ${\cal G}_s$ is
the global symmetry algebra of the supergravity sector with both electric and
magnetic dual potentials treated on equal footing \cite{w4,hidden,mtheory}. It is the extended
symmetry algebra that will determine the precise form of both the matrix
Lagrangian, as well as the particular large $N$ limits of interest \cite{mtheory},
an observation first made by us in \cite{mat1}. 

\section{The Electric-Magnetic Dual Global Symmetry Algebra}

\vskip 0.1in In our earliest proposal of a matrix formulation of a
theory of emergent spacetime \cite{mat1}, we alluded to the existence of a hidden
symmetry algebra in the matrix Lagrangian that was larger than the
obvious $U(N)$$\times$$\rm G$. A particularly obvious
indication of this fact was an $SL(10,{\bf R})$ symmetry in our
matrix Lagrangian, but we suspected that ${\cal G}_s$ was much 
larger than $SL(10;{\bf R})$ \cite{mat1}. In particular, as the reader can
guess from our discussion
in the introduction, we would like ${\cal G}_s$ to reflect the electric-magnetic 
duality of the pform gauge potentials in the supergravity sector. Of course,
thus far we have only invoked the evidence from string perturbation theory
for electric-magnetic duality in the
Ramond-Ramond sector of the type II superstrings. Based on various disparate
pieces of evidence from duality symmetries of the low energy field theory limit, 
should we not extend our conjecture to cover the pform potentials of the Neveu-Schwarz
sector? 
And what is the evidence for electric-magnetic duality in the Yang-Mills 
gauge sector? All of these questions have bearing on the precise form of
 the extended symmetry algebra of the matrix Lagrangian,  
${\cal G}_s$$\times$$\rm G$$\times$$U(N)$ .\footnote{The anomaly-free ten-dimensional
superstring theories are based on simply-laced gauge groups, which are 
electric-magnetic self-dual. But non-simply-laced groups do occur in the 
moduli spaces of CHL orbifolds \cite{cp,chl,chltalk}, namely, 
supersymmetry preserving orbifolds of
any perturbatively renormalizable and ultraviolet finite string compactification, and 
so $\rm G$ should, strictly speaking,
be replaced by $\rm G$$\times$$\rm G^*$, where $\rm G^*$ denotes the
dual magnetic gauge group \cite{gno,chl,cp}. For simply-laced groups, $\rm G^*$ 
coincides with $\rm G$.}

\vskip 0.1in From the perspective of Eguchi-Kawai planar 
reduction \cite{ek}, the $SL(D;{\bf R})$  invariance of the zero-dimensional
matrix Lagrangian in \cite{mat1} can be immediately identified as a 
remnant of the $D$-dimensional Lorentz invariance following the reduction 
of a higher dimensional field theory 
to a spacetime point.  In \cite{hidden,mtheory}, we realized that the 
field theory Lagrangians from which the class of matrix models to which
 \cite{mat1} belongs are descended via some spacetime reduction prescritpion
 must have a huge 
global symmetry algebra.  It turns out \cite{mtheory}
that they are characterized by a $U(N)$ flavor symmetry, 
a finite
dimensional Yang-Mills gauge symmetry, as well as the hidden   
Cremmer-Julia symmetry of some higher dimensional supergravity
theory \cite{cjlp}. 
The extended symmetry algebra of such matrix
Lagrangians, and their derivation from a modification of the Eguchi-Kawai 
prescription for spacetime reduction from higher dimensional field
theories, was explored in the recent paper \cite{mtheory}. We should
emphasize 
that the higher dimensional field theory Lagrangian with its $U(N)$
flavor symmetry is of no intrinsic physics interest to us; it is simply a 
convenient starting point from which to derive the matrix Lagrangian 
we desire. In particular, our prescription for spacetime reduction is not 
reversible;  the myriad continuum field theories which will
emerge from the 
double, or multiple, scaled large $N$ limits of the zero-dimensional 
matrix Lagrangian are the physical theories of interest, and the 
unphysical $U(N)$ flavor symmetry is thereby erased.

\vskip 0.1in What form does ${\cal G}_s$ take for the matrix Lagrangian
with sixteen supercharges considered in \cite{mat1}? 
The significance of electric-magnetic duality in the
appearance of the global symmetry algebras belonging to the Cremmer-Julia
sequence has been studied in a series of recent works \cite{cjlp,lp,cjlp2,w4}; 
a review appears in \cite{hidden}.\footnote{A different, and very
beautiful, direction of research on electric-magnetic
duality has explored the relationship to K theory, and to theories with
generalized cohomology
\cite{mw,sati}. The algebra of BPS states has also been explored in
\cite{mh}.} In particular, as has been clearly
elucidated by West \cite{w4}, and in work with 
Schnakenburg \cite{w1,w2,w5}, the mapping of the global symmetry algebras
of each of the ten, and eleven,
dimensional supergravities with 32 supercharges 
{\em to the single rank
eleven Lorentzian Kac-Moody algebra, ${\cal E}_{11}$,} relies 
upon the inclusion of both electric and magnetic potentials among
the generators of the algebra. Thus, without manifest electric-magnetic duality,
this remarkable algebraic unification of the theories with 32 supercharges would not hold.

\vskip 0.1in  ${\cal E}_{11}$ is, of course, the termination of the $\{ {\cal E}_{11-n} \}$ 
Cremmer-Julia
sequence of hidden symmetry groups characterizing the dimensional reductions of
11d supergravity to $n$ dimensions \cite{cjlp}, with $n$$=$$0$, and the result is no longer
a field theory, but a {\em zero-dimensional} matrix model. In practice, the procedure of
field-theoretic dimensional reduction is ill-defined in two dimensions and below.\footnote{For
a review of the subtleties which originate in the ambiguity in dualizing a gauge
field in less than three dimensions, we refer the reader to the discussions in  
\cite{lp,cjlp,cjlp2,hidden}, which
include citations to the original literature.} We should emphasize that West's
recent identification of ${\cal E}_{11}$ as the symmetry algebra of theories with 32
supercharges in \cite{w4} does not rely on dimensional reduction, nor on the usual
dualization of dimensionally-reduced fields followed by a mapping to a non-linear 
realization in order to
identify the exponentiated group element as belonging to some known algebra.
This traditional methodology \cite{cj,cjlp} would not work anyhow for reductions to 
below three dimensions. West's new methodology for uncovering an ${\cal E}_{11}$
symmetry in theories with 32 supercharges is reviewed by us in \cite{hidden}. We 
will apply it in the context of theories with 16 supercharges in what follows below.

\vskip 0.1in 
Neither can the reduction of a higher dimensional field theory to a spacetime point
a la Eguchi-Kawai result in a zero-dimensional matrix model preserving the 
symmetries of the Einstein
supergravities, a distinction clarified in our recent paper \cite{mtheory}.
Two significant changes are required in order to derive the zero-dimensional 
matrix Lagrangian proposed by us in \cite{mat1} from a higher dimensional field
theory by spacetime reduction. First, we must
begin with a higher-dimensional supergravity-Yang-Mills Lagrangian with large 
$N$ {\em flavor} symmetry. It is essential that the gravitational zehnbein
and the Yang-Mills potential live in identical $N$$\times$$N$ representations of 
the $U(N)$, thereby ensuring that 
the extended symmetry algebra of the matrix Lagrangian takes the direct 
product form: $U(N)$$\times$$\rm G$$\times$${\cal G}_s$.  Supersymmetry 
commutes with this algebra, with the fermionic superpartners living in identical
$U(N)$ representations. Note that this is also the global 
symmetry algebra of the higher dimensional field theory, except that 
there we
can distinguish $U(N)$ as a flavor symmetry,
$\rm G$ as a Yang-Mills gauge symmetry, and ${\cal G}_s$ 
as the global symmetry algebra of
the supergravity sector.
Next, we must modify the Eguchi-Kawai prescription which simply drops all
space and time derivatives in the reduction of the field theory
to a single point in spacetime. In our spacetime reduction prescription, we
instead replace all fields by linearized Taylor expansions about the origin of the
local tangent space at a single point in the spacetime manifold. This preserves,
in particular, the notion of a first order partial derivative in the local
tangent space, allowing us to assign $U(N)$ matrix-valued definitions to both
the spacetime coordinates, and spacetime derivatives, consistent with the basic
rules of differential geometry. The $N$$\times$$N$ matrix variables,
$\{ E_{\mu}^a , E_a^{\mu} ; \phi \}$, with $\mu$, $a$ $=$ $0$, $\cdots$, $9$,
encapsulate the basic information on the 
emergent spacetime geometries. In the large $N$ limit, as many as ten, or 
eleven, upon including the dilaton scalar mode, noncompact coordinates, 
and coordinate derivatives, can emerge from the diagonalized eigenvalue 
configurations of this basis of matrix variables. Details can be found in the
references \cite{mat1,hidden,mtheory}. 

\vskip 0.1in We emphasize that without a prescription
to build an emergent {\em coordinate derivative} in the large $N$ limit, it is 
not possible for the full ${\cal G}_s$ supergravity symmetry to emerge from 
a given matrix Lagrangian. Nor would we have the requisite 
tools to build up emergent
curved spacetime geometries in the large $N$ limit. The difficulties in 
reconstructing the full nonlinear gravitational interaction of the Einstein
action, beyond the nonrelativistic Newtonian potential, and of
incorporating curved spacetime backgrounds in addition to flat 
spacetime, were two of the major obstacles to making progress on the
M(atrix) theory conjecture \cite{bfss,ikkt,plefka}. 

\vskip 0.1in What is the precise form of ${\cal G}_s$ for the matrix Lagrangian
proposed in \cite{mat1,mtheory}? The question can be addressed by using
 a recent work of West and Schnakenburg \cite{w4}, where the global 
symmetry algebra of the $N$$=$$1$ d=10 chiral supergravity has been
analyzed. Recall that the
10d chiral supergravity with sixteen supercharges is, a priori, an anomalous
field theory, but it can be self-consistently coupled
to Yang-Mills gauge fields in order to achieve anomaly cancellation. 
The gauge, gravitational, and mixed, 
anomaly-free choices of gauge group are $E_8$$\times$$E_8$ and $SO(32)$, and
these are the 10d supergravity-Yang-Mills theories appearing in the low
 energy limits of the heterotic and type I superstring
theories \cite{polbook}. The matrix Lagrangian in \cite{mat1}
was proposed by us as a {\em nonperturbative} formulation of 
these perturbatively renormalizable
and anomaly-free superstring theories.

\vskip 0.1in 
Restricting to the fields in the 10d $N$$=$$1$ chiral supergravity 
sector alone, and using the notation of \cite{w4}, we have the following 
symmetry generators:
\begin{equation}
K^a_b , R , R^{c_1 c_2} , R^{c_1 \cdots c_6} , R^{c_1 \cdots c_8}
 \quad .
\label{eq:typei}
\end{equation}
The supergravity sector contains a zero-form dilaton and an antisymmetric
two-form potential, in addition to their 10d Poincare-Hodge duals:
respectively, the magnetic dual eight-form, and six-form, gauge potentials. The
$K^a_b$ are the generators of $GL(10,{\bf R})$; linear combinations of these
generators can be shown to span the 10d Lorentz algebra 
\cite{w4}. The commutator algebra of the generators in Eq.\ (\ref{eq:typei}) 
was analyzed in
\cite{w5}. It takes the form: 
\begin{equation}
[K^a_b , K^c_d ] = \delta_b^c K^a_d
- \delta_d^a K^c_b , \quad\quad [K^a_b , P_c ] = \delta_c^a P_b ,
\quad\quad [K^a_b , R^{c_1 \cdots c_p } ] = \delta_b^{c_1} R^{ac_2
\cdots c_p} + \cdots  \quad , \label{eq:gl10}
\end{equation}
plus the simplified algebra of 0, 2, 6, and 8-form generators:
\begin{equation}
[R , R^{c_1 \cdots c_p} ] = d_p R^{c_1 \cdots c_p}  , \quad
[R^{c_1 \cdots c_p} , R^{c_1 \cdots c_q} ] = c_{p,q} R^{c_1 \cdots
c_{p+q}}
 \quad . \label{eq:extrabI}
\end{equation}
The numerical values of the coefficients can be computed directly
by requiring self-consistency with the Jacobi identities, and the equations 
of motion,
as shown in \cite{w4}. A useful self-consistency check on this algebra
was performed
by us in \cite{hidden}. We showed that the coefficients obtained in 
\cite{w4} agree  
precisely
with those inferred from a chirality 
projection on the global symmetry algebra of the type IIB
chiral 10d supergravity obtained in \cite{w2}.
Setting to zero the extra forms removed by the chirality projection 
in Eqs.\ (1.1-1.3) of the latter reference, we found that
the remnant non-vanishing structure constants take the simple
form \cite{hidden}:
\begin{equation}
d_{q+1} = - \quart (q-3) ,  ~ q=1,5, \quad\quad \quad c_{2,6} =
\half \quad . \label{eq:strbI}
\end{equation}
The self-consistency of the global symmetry algebras
is a low energy reflection of the well-known connection between
the 10d type IB, and type IIB, superstring theories, following the orientation projection 
to the symmetric combination of 
 left-moving and right-moving modes on the worldsheet of the
 type IIB superstring 
\cite{polbook}: the orientifold plane breaks half the supersymmetries
of the type IIB theory, giving a chiral $N$$=$$1$ supergravity with 
graviton-dilaton multiplet, and an antisymmetric two-form potential in the
Ramond-Ramond sector. Up to field redefinitions, and rescalings of the
couplings, this supergravity matches with that obtained in the low energy limit of the 10d
heterotic superstring. It should be noted that the two-form potential now
appears, instead,
in the Neveu-Schwarz
sector \cite{polbook}. Finally, unlike the heterotic string, the type IB string can
accommodate additional Ramond sector background fields, but a precise
correspondence can be found between the mass spectrum and couplings 
of the two superstring theories in all of the backgrounds with sixteen supercharges 
\cite{flux}, as we will describe later in this paper.

\vskip 0.1in How does the global
symmetry algebra of the 10d heterotic-type I chiral supergravity, ${\cal G}_{\rm IB}$, 
relate to the symmetry algebra of theories with 32 supercharges? From our
demonstration above, it is clear that ${\cal G}_{\rm IB}$ $\in$ ${\cal G}_{\rm IIB}$. 
West has shown \cite{w4} that the global symmetry
algebras of {\em each} of the ten, and eleven, dimensional supergravities
with 32 supercharges can be mapped to the single rank eleven Lorentzian Kac-Moody
algebra ${\cal E}_{11}$. This algebra is also known as the rank 11 very-extension of
the compact Lie algebra $E_8$, denoted as 
$E^{+++}_{8}$$=$$E_8^{(3)}$, where the superscript signifies the extension of the Dynkin diagram
of $E_8$ by three nodes: affine, over, and very.  The explicit construction, and some simpler
aspects of the representation theory, of very-extended algebras can be found in \cite{gow}. 
If we compare the generators
and commutation rules of ${\cal G}_{\rm IB}$ with the standard Chevalley basis for the
very-extended algebra $E_8^{(3)}$, written in either its IIA or IIB guises as
shown in \cite{w4}, we find that we are missing some of the
positive root generators in either formulation. We have all of the
generators, $E_a$$=$$K^a_{a+1}$, $a$$=$$1$, $\cdots$, $9$, of
$GL(10,{\bf R})$. In the IIA formulation, given in Eq.\ (4.4) of
\cite{w4}, we are missing the roots corresponding to the R-R
one-form, and NS-NS twoform, namely, $E_{10}$$=$$R^{10}$, and
$E_{11}$$=$$R^{910}$. In the IIB formulation, we are missing the
roots labelled $E_{9}$$=$$R_1^{910}$, and $E_{10}$$=$$R_2$,
arising, respectively, from the NS-NS two-form potential, and R-R
scalar. It is clear we cannot build a full ${\cal E}_{11}$$=$$E_{8}^{(3)}$ algebra
from the restricted set of generators in ${\cal G}_{\rm IB}$.

\vskip 0.1in In \cite{w5}, it was pointed out that a different
rank eleven very-extended algebra, namely, the very-extension of
the $D_8$ compact subalgebra of $E_8$, can be spanned by the generators of
${\cal G}_{\rm IB}$. The authors of \cite{w5} identify the isomorphism 
$E_a$$=$$K^a_{a+1}$, $a$$=$$1$, $\cdots$, $9$, $E_{10}
$$=$$R^{910}$, and $E_{11}$$=$$R^{5678910}$. This choice of simple roots
can be shown to generate the very-extended algebra $D_8^{(3)}$.
It should
be noted that the two-form and six-form potentials are
Poincare-Hodge duals in ten dimensions, and the
 isomorphism identified in \cite{w5} 
includes {\em both} in the simple root basis for ${\cal D}_{11}$.  It is interesting
that this differs in
spirit from the isomorphisms 
mapping, respectively, ${\cal G}_{\rm IIA}$,  ${\cal G}_{\rm mIIA}$, ${\cal G}_{\rm IIB}$, or
${\cal G}_{11}$, to the Chevalley basis of $E_{8}^{(3)}$ \cite{w4}; neither employs
a {\em pair} of electric and magnetic dual potentials within the simple root basis. 
Moreover, the 
generators of the rank eleven very-extension
of $D_8$ appears to encapsulate the basic physical content of 
the supergravity sector of  
renormalizable superstring theories remarkably well: 
the graviton, the scalar dilaton and antisymmetric two-form potentials, plus their
10D Poincare-Hodge duals. We find the isomorphism of
${\cal G}_{\rm IB}$ to the rank eleven algebra ${\cal D}_{11}$, discovered in
section 1 of  \cite{w5}, to be remarkably compelling, and also 
aesthetically pleasing.

\vskip 0.1in 
The 10d $N$$=$$1$ chiral supergravity is, of course, anomalous, and must be
extended by a Yang-Mills sector with suuitable nonabelian gauge group in order
to achieve consistency as a quantum theory. This requires inclusion of the 
massive Kalb-Ramond term
in the 10d effective action, at leading order in $m_s^{-1}$ \cite{br,polbook}. 
Including the full 
slew of  $O(m_s^{-1})$ corrections present 
in the target spacetime Lagrangian of the 
heterotic, or type I, superstring theories, where the mass scale,
$m_s = \alpha^{\prime -1/2}$, is the tension of the fundamental
closed string \cite{polbook}, implements the beautiful phenomenon of 
perturbative renormalizability in these theories. The precise form of the target spacetime
Lagrangian of the heterotic string theory, inclusive of terms up to fourth
order in the $\alpha^{\prime}$ expansion, has been derived by Bergshoeff
and de Roo in \cite{br}. What is the global symmetry algebra of the leading
terms in the $\alpha^{\prime} $ expansion of this Lagrangian? The 
leading terms correspond
to those in the 10d $N$$=$$1$ Einstein-Yang-Mills Lagrangian, and these 
can be 
obtained by the Noether method \cite{fs}, checking invariance under
local Lorentz and Yang-Mills gauge transformations, in addition to local 
supersymmetry transformations. The spacetime Lagrangian of the perturbative ten-dimensional
$N$$=$$1$ superstring theories has been shown to have
dual, two-form and six-form,  
formulations \cite{br}, evidence for at least a partial electric-magnetic duality
in the {\em nonperturbative} superstring theory which shares the same low energy
limit. In addition, the 
type IB--heterotic strong-weak coupling duality conjecture maps the D5branes of the
type I string to the NS5branes of the heterotic string \cite{witsm,gimp}. Thus, string ground
states with eight supercharges have given significant evidence and physical insight into
the dynamics of heterotic fivebranes.
Notice that the perturbative mass spectrum of the heterotic string 
only contains the massless scalar dilaton and 
antisymmetric two-tensor potential, both arising in the Neveu-Schwarz sector.  
From the perspective of the type IB superstring, the dilaton arises in the Neveu-Schwarz
sector, while the antisymmetric two-form is in the Ramond sector. 

\vskip 0.1in 
Thus, although the perturbative string mass spectrum has only given us the electric 
potentials of this theory,  a full electric-magnetic duality in the spectrum of supergravity 
potentials in the nonperturbative theory would imply the global symmetry algebra 
${\cal D}_{11}$$\times$$\rm G$, where $\rm G$ is the anomaly-free Yang-Mills group,
respectively, $E_8$$\times$$E_8$ or $SO(32)$. There is at least partial
evidence that the spacetime Lagrangian of this theory admits both electric, and dual
magnetic, formulations. Passing between the twoform and sixform formulations involves 
a simple set of field 
redefinitions, and relations between couplings \cite{br,fs}, and preliminary evidence
has been given for a dual graviton formulation of the kinetic terms in the bosonic sector 
of 11d supergravity \cite{w1}.
Thus, we can conjecture 
that the extended symmetry algebra underlying the matrix formulation for the
nonperturbative type I-I$^{\prime}$-heterotic string theories in \cite{mat1,mtheory} 
takes the form 
${\cal D}_{11}$$\times$$\rm G$$\times$$U(N)$. In the next section, we will 
see how the emergent spacetime backgrounds accessible as large $N$ limits of the
matrix Lagrangian in \cite{mat1,mtheory} gives rise to a large class of type I, type II,
and heterotic string backgrounds in diverse spacetime dimensions, $d$ $\le$ $10$,
all of which preserve sixteen supercharges, and also 
correspond to perturbatively renormalizable and ultraviolet finite 
superstring theories. Our discussion
will give rise to a natural question one could frame much more broadly: should a
nonperturbative formulation for String/M theory be electric-magnetic duality covariant, 
or merely duality invariant?

\section{Large N Limits \& the String/M Duality Web}

\vskip 0.1in The nature of the isomorphism pointed out by us in \cite{mat1,mtheory}
between large N limits of the matrix Lagrangian, and 
weak coupling limits of the String/M Duality web, is 
as follows. The matrix Lagrangian 
in \cite{mat1} has the global symmetries of
the 10d $N$ $=$ $1$ chiral supergravity-Yang-Mills theory \cite{br,fs}, which we will
denote as ${\cal G}_s$$\times$$\rm G$,
in addition to the obvious $U(N)$ flavor symmetry of 
the generic $U(N)$ matrix model. 
As described in
section 2, ${\cal G}_s$$\times$$\rm G$ can plausibly
be made as large as ${\cal D}_{11}$$\times$$\rm G$, depending on how much of the electric-magnetic
duality can be made manifest in the Lagrangian itself. Given the range of
potentials represented in the matrix Lagrangian, the large $N$ limit allows for 
many inequivalent scalings, generalizing the two-parameter, $(g_s  , N)$,
double-scaling limit introduced for the $c$ $=$ $1$ matrix models \cite{mm}. For the
$U(N)$ matrix Lagrangian with sixteen supersymmetries in \cite{mat1,mtheory},
written in its \lq\lq heterotic" two-form 
formulation, we 
have the following list of parameters
to work with:
\begin{equation}
(g_s =e^{{\bar{\phi}}},  ~ {\bar{E}}^{\mu}_a  ,  ~  {\bar{E}}^{a}_{\mu} ,  ~ {\bar{A}}_{a}, ~  {\bar{B}}_{ab}).  \quad 
({\bar{F}}_{ab}, ~ {\bar{H}}_{abc}).   \quad \quad  ~( {\bar{B}}_{c_1 \cdots c_p}, {\bar{H}}_{c_1 \cdots c_{p=1}}), ~
p=6 ,  8 \quad .
\label{eq:list}
\end{equation}
We have grouped together the {\em electric potentials}, namely, the scalar
dilaton, graviton, antisymmetric two-form potential, and Yang-Mills vector potential, separate
from the {\em electric field strengths}, thereby allowing for the possibility of both background
fields, and 
background fluxes. Finally, there are the {\em magnetic
potentials}, the dual eightform and sixform potentials, and their corresponding {\em magnetic
field strengths}.
Their presence in the matrix Lagrangian is
only implicit: the Lagrangian admits a duality transformation on the \lq\lq electric"
matrix variables, so we are free to consider the dual matrix Lagrangian, with a dual set of
large $N$ scalings, prior to taking the large $N$ limit. The existence of the dual 
six-form formulation of the continuum spacetime Lagrangian is well-known \cite{br,fs,ss}, and
some preliminary evidence has also 
been given for a dual graviton formulation of the 11d supergravity 
Lagrangian \cite{w1}. 

\vskip 0.1in
The indices in the equation above
run from $0$ to $9$. The $N$$\times$$N$ zehnbein matrix variables,
$(E^{\mu}_a$, $E^a_{\mu})$, encapsulate the data specifying the background target spacetime geometry. As many as ten continuum spacetime coordinates, and ten continuum
spacetime coordinate derivatives, can emerge in the large $N$ limit \cite{mtheory}. 
The respective scalings of 
the dilaton, and the zehnbein, with powers of $N$ and $m_s$ $=$ $\alpha^{\prime -1/2}$,
are determined by requiring the canonical overall normalization for the
10d Einstein action, as well as the relative normalization of the kinetic term for the dilaton.
Recall that the \lq\lq type IB" and \lq\lq heterotic" 
guises of the 10d N=1 spacetime Lagrangian
are distinguished by the scaling of the Yang-Mills kinetic term relative to
the Einstein kinetic term \cite{witdual,polbook}; the Yang-Mills gauge fields 
appear in the open string sector in type IB, and $g_{s} $  $=$ $ g_{\rm open}^2$. 
In addition, the antisymmetric two-form
potential appears in the Ramond sector of type IB, and the 
kinetic terms for Ramond sector fields do not contain powers of $[e^{\phi}]$ \cite{witdual,polbook}. 
These two differences will
determine the relative scalings for $A_a$, $B_{ab}$ in the large
$N$ limits that distinguish type IB from heterotic.
Scaling as would be
appropriate for the 10d heterotic 
Lagrangian, implies one of  
two anomaly-free choices for the gauge group, $SO(32)$ or $E_8$$\times$$E_8$. For the type IB scaling, we must
instead choose the gauge group $SO(32)$. 

\vskip 0.1in  Let us move on to explaining
the emergence of the toroidal backgrounds.
Suppose we take the large $N$ limit of the matrix Lagrangian, scaling $(g_s$,
$E^{\mu}_a$, $E^a_{\mu})$ with the appropriate powers of $N$, $m_s$ to give the correctly normalized Einstein spacetime action 
in dimensions $d$ $<$ $10$. For a toroidal compactification, the 
$({\bar{E}}^{\mu}_a$, ${\bar{E}}^a_{\mu})$ will decompose
naturally into noncompact, and compact, coordinates and coordinate derivatives. Notice that the
decomposition of the remaining matrix variables into components carrying a noncompact, or compact,
index, proceeds in precise analogy with how one constructs the 
dimensional reductions of the fields in a continuum spacetime Lagrangian \cite{witdual,cjlp}. 
In particular, the T-duality symmetries that
characterize the moduli space for a given compactification become manifest: associating with 
each compactification its T-duality group, we can identify the isomorphism between 
large $N$ limits and moduli spaces. Finally, this prescription
carries over naturally to the supersymmetry preserving heterotic  
CHL orbifolds \cite{cp,cl,chltalk}: the orbifold projection on the 
string mass spectrum identifies a specific truncation of the moduli space of massless scalars, 
also isomorphic to a corresponding truncation of the finite $N$ matrix Lagrangian. 
To obtain the
string effective Lagrangians that describe 9D 
CHL orbifolds of the type IIA, or type I$^{\prime}$, string theories, preserving only 
16 supersymmetries \cite{cl}, we must incorporate the Ramond sector one-form 
potential, ${\bar{C}}_a$, in the definition of a distinct, and nontrivial, large $N$ scaling limit.

\vskip 0.1in More generally, notice that the inequivalent multiple-scaled 
large $N$ limits of the matrix Lagrangian are
distinguished by the individual scalings of
{\em any} combination of the background parameters with appropriate powers of
$N$, the string mass-scale, $m_s =\alpha^{\prime -1/2}$, and the dimensionless string
coupling, $g_s =e^{{\bar{\phi}}}$, 
while taking $N$$\to$$\infty$. 
One final comment: if $\rm G$ is 
a non-simply-laced group, as happens at certain enhanced symmetry points 
in the CHL moduli spaces, we must include 
the duality-transformed vector potential of the magnetic dual group, $A^*$, among 
the magnetic
potentials listed in Eq.\ (\ref{eq:list}).

\vskip 0.1in
Setting $\alpha^{\prime}$ to zero in the multiple-scaled large N matrix Lagrangian 
recovers the massless low energy 
field theory limit of some perturbatively renormalizable string ground state: this 
is by construction, since the choice of
matrix Lagrangian was determined by its global symmetries.
We have conjectured 
that, as a consequence,  {\em the off-diagonal $O(1/N)$ terms in the
matrix Lagrangian
for a particular choice of large $N$ scaling will be isomorphic to the 
$O(\alpha^{\prime})$ corrections
in a corresponding perturbatively renormalizable ground state of string theory
\cite{mtheory}}.
To anybody familiar with calculating the string massive mode spectrum, mass-level
by mass-level, this conjecture should come as no surprise: the primary fields of
the worldsheet 
conformal field theory appear in the low energy massless field theory limit, and 
their descendants
fill in the tower of massive string modes. Thus, the algebraic symmetry structure
that determines 
the massless field theory limit will also be replicated at each mass level.\footnote{The
symmetries of the string mass spectrum and perturbative S-matrix have been the focus of
several previous works in string theory \cite{bgm,zuk,latt}. Related work on the 
high-energy symmetries of String Theory appears in \cite{gross,witz,borch,moore}.} This
property is characteristic of the $O(1/ N)$ corrections to the large $N$ limit
of any matrix Lagrangian. 

\section{Conclusions}

\vskip 0.1in The Lagrangian in \cite{mat1} was originally proposed by us
as a matrix 
formulation for nonperturbative type I-heterotic string theory. It was argued that
spacetime 
itself, and the target spacetime Lagrangians of perturbatively renormalizable 
string theories with sixteen supercharges, emerge from the 
large $N$ limits of this matrix Lagrangian \cite{mtheory}. The matrix 
Lagrangian in \cite{mat1} shares the global symmetries of the chiral 10d N=1 
supergravity-Yang-Mills theory, in addition to the obvious $U(N)$ symmetry of a 
generic matrix model.  

\vskip 0.1in The significance of the global symmetry algebra in determining 
what aspects of the electric-magnetic duality symmetries of string/M theory 
backgrounds in diverse spacetime dimensions should be made {\em manifest} in the
matrix Lagrangian itself, is the general theme explored in sections 1 thru 3 of this
paper. We reviewed the evidence from the worldsheet formalism of perturbative string theory 
for the eleven-dimensional electric-magnetic duality in the Ramond-Ramond
sector of the type II superstring theories and Dpbrane spectrum \cite{dbrane,flux,hodge}. 
If both electric
and magnetic dual generators are included in the global symmetry algebra on
equal footing, then, as has been 
shown by West and collaborators \cite{w4,w5}, the algebra extends
to a rank eleven Lorentzian Kac-Moody algebra: ${\cal E}_{11}$ for theories
with 32 supercharges, and ${\cal D}_{11}$ for the theories with sixteen 
supercharges. It should be noted, as was pointed out in my 1995 paper
with Polchinski \cite{cp,chltalk}, that electric-magnetic
duality in theories with sixteen supercharges only
becomes {\em manifest} in the {\em four-dimensional} moduli spaces, in the guise of
S-duality \cite{br,ss,cp}. Innovative formalisms have been
developed for implementing fully geometric realizations of self-duality \cite{cjlp2,hullg},
but no completely duality covariant formulation exists for any of the supergravities,
or string theories, at the present time. 
The precise role of the underlying eleven-dimensional
electric-magnetic duality as it operates within the String/M Duality web, the 
associated algebras, and the question of whether we need a duality {\em covariant},
or merely a duality {\em invariant}, formulation for nonperturbative string/M theory, 
therefore remain questions that will need further exploration.

\vskip 0.1in The matrix proposal for nonperturbative String/M theory given by us in 
\cite{mat1,mtheory}
has been modestly successful in establishing
self-consistency with the type I-heterotic-M strong-weak coupling duality conjectures,
in a multitude of disconnected moduli spaces, and in diverse spacetime dimensions. The 
nature of the disconnectedness of string moduli spaces, and their physical interpretation, 
has been described in our recent work \cite{chltalk}.
Our discussion in section 3 of the myriad inequivalent large N 
scaling limits of a single finite N matrix Lagrangian
clarifies that the existence of isolated low
energy Universes in string theory does not, in itself,
imply an essential role for the {\em Anthropic 
Principle} \cite{ap,bp,denef} in quantum cosmology. At leading 
order in the $\alpha^{\prime}$ expansion, the efficacy of this result is self-evident.
We have conjectured further that the $1/N$ expansion
of the matrix Lagrangian, for any given choice of scalings, 
provides the systematics of the $\alpha^{\prime}$
corrections to the massless low energy field theory limit in some 
renormalizable string ground state \cite{mtheory}. A confirmation of
this conjecture in the perturbatively renormalizable flat spacetime backgrounds
would be a significant step forward in establishing the
viability and uniqueness of our proposal for nonperturbative String/M theory.

\vskip 0.1in
Conventional methods for computing
the perturbative superstring S-matrix, thereby inferring higher order terms in
the target spacetime Lagrangian, although 
straightforward in principle have been cumbersome in practice 
\cite{gsloan,polbook}. Thus, even in 
exactly solvable flat spacetime backgrounds where the worldsheet formalism
is extremely well-developed, it is often difficult to write down precise expressions
for terms at high order in the $\alpha^{\prime}$ expansion. The self-consistency 
of an exactly solvable 
worldsheet conformal field theory description, when it exists, suffices to establish the 
{\em existence} of the corresponding perturbatively renormalizable, all-orders-in-$\alpha^{\prime}$,  
spacetime Lagrangian. It should be kept in mind that, if the supersymmetry 
breaking scale in Nature is far below the string mass scale, the higher $O(1/m_s)$ 
corrections are in any case irrelevant for the
purposes of low energy physics in such backgrounds \cite{chltalk}. But at intermediate values of the 
string coupling constant, or in genuinely nonperturbative
backgrounds of string theory where we do {\em not} have an exactly 
solvable worldsheet formulation, 
the tantalizing possibility that we might nevertheless be able to compute the higher
order corrections in $\alpha^{\prime}$, thus also establishing renormalizablity, is
intriguing, to say the least. 

\vskip 0.2in \noindent{\bf ACKNOWLEDGMENTS}

\vskip 0.1in I am grateful to Lisa Carbone and Greg Moore for the invitation to present
this work. I would like to thank Paul Aspinwall, Lisa Carbone, 
Bernard De Wit, Jim Hartle, 
Hikaru Kawai, Costas Kounnas, Emil Martinec, Greg Moore, 
Herman Nicolai, Sanjaye Ramgoolam, 
Hisham Sati, John Schwarz, Savdeep Sethi, Steve Shenker, and Peter West, for 
enjoyable discussions on some of these ideas.

\end{document}